\documentclass[graybox]{svmult}

\usepackage{mathptmx}       
\usepackage{helvet}         
\usepackage{courier}        
\usepackage{type1cm}        
\usepackage{makeidx}         
\usepackage{graphicx}        
\usepackage{epstopdf}
\usepackage{multicol}        
\usepackage[bottom]{footmisc}

\usepackage{amsmath}

\makeindex             

\begin{document}

\title*{Local stability conditions and calibrating procedure for new car-following models used in driving simulators}
\author{Valentina Kurc and Igor Anufriev}
\institute{Valentina Kurtc \at St. Petersburg State Polytechnical University, Polytechnicheskaya 29, 195251 St. Petersburg, Russia, \email{kurtsvv@gmail.com}
\and Igor Anufriev \at St. Petersburg State Polytechnical University, Polytechnicheskaya 29, 195251 St. Petersburg, Russia \email{igevan@mail.ru}}
%
%
\maketitle

\abstract{The Intelligent Driver Model (IDM) is studied and several drawbacks with respect to driving simulators are defined. We present two modifications of the IDM. The first one gives any predefined distance to the leading vehicle in a steady state. The second modification is a combination of the first one and the optimal velocity model. It takes into account driver's reaction time explicitly and is described by delay differential equation. This model always results in realistic vehicles accelerations what allows simulating real traffic collisions. \newline\indent
Necessary and sufficient conditions are obtained, that guarantee a non-oscillating solution near the equilibrium for the vehicle platoon. We suggest the calibrating framework based on a numerical solution of the constrained optimization problem. Nonlinear constraints are generated by the numerical integration scheme. The suggested procedure incorporates the local stability conditions obtained and takes into account vehicle dynamics, drivers' behavior and weather conditions.}
\section{Introduction}
\label{sec:1}
{Driving simulators are certain kind of training systems in a car driving application \cite{Allen, Roberts}. Firstly, these were developed for the training in the use of military mechanisms during the Second World War. Later, driving simulators were used to examine drivers' behavior and their interaction with the environment - vehicle controls, other cars, pedestrians and etc. The quality of traffic simulation plays a major role. It is important that the other cars move as naturally as possible, drivers behave in predictable manner and in accordance with traffic rules. This article focuses on appropriate vehicle traffic models for using in driving simulators. The general and necessary features which a model should demonstrate are real vehicle dynamics, adequate driver's behaviour and mathematical stability of the solution. \newline\indent
In this paper, we propose two microscopic models. The first one provides any predefined distance in a steady state that allows to take into account road covering and weather conditions. The second modification is the extension of the first one and includes driver's reaction time. Both models are described by continious acceleration functions and are the car-following type ones. \newline\indent
In Sect.~\ref{sec:2} we state some drawbacks of the Intelligent Driver Model (IDM) in application to driving simulators. Then we formulate two modifications basising on the IDM in terms of the acceleration function. In Sect.~\ref{sec:3} we evaluate linear stability analysis for the case of vehicle platoon for the first proposed model and obtain condition for non-oscillating solution. In Sect.~\ref{sec:4} we present calibration framework which is applicable for any microscopic model. It allows considering local stability conditions obtained. In the concluding Sect.~\ref{sec:5}, we discuss results and further investigations.
}
\section{IDM drawbacks and two modifications}
\label{sec:2}
{The trainee observes vehicle traffic from their respective cabin so the cars' dynamics and drivers' behavior should be realistic to the highest degree. We considered the Intelligent Driver Model (IDM) \cite{Treiber} as a starting point for using in driving simulators. This model has a set of parameters which can be tuned to achieve desired objectives. After detailed study of this model we discovered several disadvantages in application to driving simulators.
\begin{itemize}
	\item{The IDM admits big deceleration values \cite{Kesting2010} and, as a result, a collective dynamics is crash-free. However, accidents are probable when people study driving.}
	\item{Instant reaction to the leading car. For example, at the signalised intersection cars start simultaneously.} 
	\item{The steady-state gap does not incorporate the road’s covering effect
	\begin{equation}
		d^*_{IDM}(v) = (s'+Tv)/{\sqrt{(1+(v/v^0)^\delta)}}
	\label{eq:01}
	\end{equation}}
\end{itemize}
To eliminate these drawbacks two modifications are presented.}

\subsection{Predefined distance in a steady state}
\label{subsec:1}
{Let us formulate the acceleration function for the first proposed model
\begin{equation}
	\dot{v}=w(h,d^*,D)a\left(1-\left({v}/{v^0}\right)^\delta\right)+(1-w(h,d^*,D))a\left(1-\left({d^*}/{h}\right)^2\right)
\label{eq:02}
\end{equation}
Here $w$ is a continuous weight-function, which depends on headway $h$ and parameter $D$.
\begin{equation}
w(h,d^*,D) = \left\{ 
  \begin{array}{l l}
    0, h \in (-\infty, d^*) \\
   -2\left(\frac{h-d^*}{D}-1\right)^3-3\left(\frac{h-d^*}{D}-1\right)^2+1, h \in [d^*, d^*+D] \\
		1, h \in (d^*+D, +\infty)
  \end{array} \right.
\label{eq:03}
\end{equation}
If vehicle is quite far from its leader Eq.~\ref{eq:02} has the first term only. If the current gap is less than the steady-state distance set, only the second term works.  \newline\indent
This car-following model guarantees any predefined gap between vehicles in the steady-state flow. Equalling the right-hand side of Eq.~\ref{eq:02} to zero, it is not difficult to demonstrate that the distance in a steady state is $d^*$. We chose distance from the Tanaka model \cite{Gartner}. It contains ‘the braking path’ and the coefficient $c$ that characterizes the road’s covering
\begin{equation}
d^*(v) = s'+Tv + cv^2\;,
\label{eq:04}
\end{equation}
}
\subsection{Reaction time}
\label{subsec:2}
{To take into account driver's reaction time we suggest the second modification that is desribed by the DDEs. It combines the optimal velocity model \cite{Bando} and the first modification. The acceleration function is as follows
\begin{equation}
\dot{v_i}=w(h_i,d^*,D)a\left(1-\left({v_i}/{v^0}\right)^\delta\right)+(1-w(h_i,d^*,D))b\left(V(h_i(t-\tau)) - v_i\right)\;,
\label{eq:05}
\end{equation}
Here we follow the same logic as in the first model (\ref{eq:02}) - separate free-road vehicle dynamics and interaction with its leader. Moreover, we suggest to make the parameter $s$ of the optimal velocity function \cite{Orozs} depending from the current distance to the leader $h$
\begin{equation}
V(h) = \left\{ 
  \begin{array}{l l}
    0, 0 \le h \le s' \\
		v^0\frac{\left(\frac{h-s'}{s(h)}\right)^3}{1+\left(\frac{h-s'}{s(h)}\right)^3}, h > s'
  \end{array} \right.
\label{eq:06}
\end{equation}
This empirical relationship is obtained as follows. We put the vehicle with the velocity $v$ at a distance of the Tanaka model (\ref{eq:04}) and request it to stop at the distance $s'$ from its fixed leader. \newline\indent
Fig. \ref{fig:1} compares acceleration dynamics of the IDM and the model (\ref{eq:05}). According to the model with the delay driver starts deceleration earlier and does it more smoothly with approximately constant rate.
\begin{figure}[h]
\includegraphics[width = 0.9\textwidth]{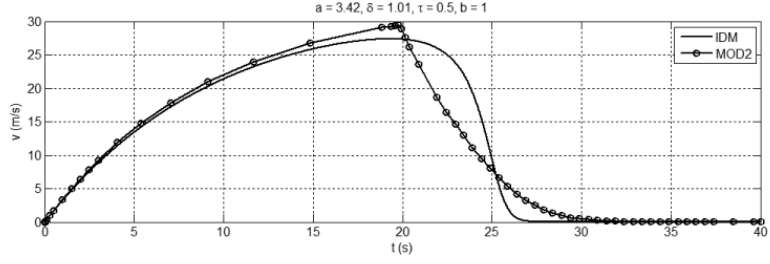}
\caption{Comparison of the IDM and model with delay proposed. Fixed leader approaching - velocity time history.}
\label{fig:1}       
\end{figure}
}

\section{Linear stability analysis}
\label{sec:3}
{Here we perform a linear stability analysis for a vehicle platoon. Lets the acceleration of the $i$th vehicle be set on the basis of its speed $v_i$, the leader’s speed $v_{i-1}$ and the distance between $h_i$
\begin{equation}
\dot{v_i}=a^{(i)}(v_{i-1}, v_i, h_i)
\label{eq:07}
\end{equation}
Every vehicle may have its own acceleration function. We have a system of $2n$ ODEs for $n$ vehicles. The Jacobian matrix has the block-like structure, so it is possible to obtain the determinant analytically
\begin{equation}
\ \chi(\lambda)=\prod_{i=0}^n (a^{(i)}_h-\lambda(a^{(i)}_v-\lambda))
\label{eq:08}
\end{equation}
Here $a^{(i)}_h$ and $a^{(i)}_v$ are partial derivatives at the steady-state point. Finally we obtain all 2n eigenvalues. The local stability takes place when its real parts of all eigenvalues are negative. If we need no oscillations we require the imaginary parts to be equal to zero:
\begin{equation}
\left\{ 
  \begin{array}{l l}
    \text{local stability, } & a_v^{(i)} < 0, \\
   \text{no oscillations, } & (a_v^{(i)})^2-4a_h^{(i)} > 0.
  \end{array} \right.
\label{eq:09}
\end{equation}

Now lets apply these conditions to the first model presented. We consider the car approaching its fixed leader. The results are shown at Fig.~\ref{fig:2}. The horizontal and vertical axes represent the velocity and the distance to the fixed leader respectively. On the left plot the parameters' values do not sutisfy inequality obtained and oscillations over the equilibrium take place, whereas the right plot represents their absence.
\begin{figure}[h]
\includegraphics[width = 1\textwidth]{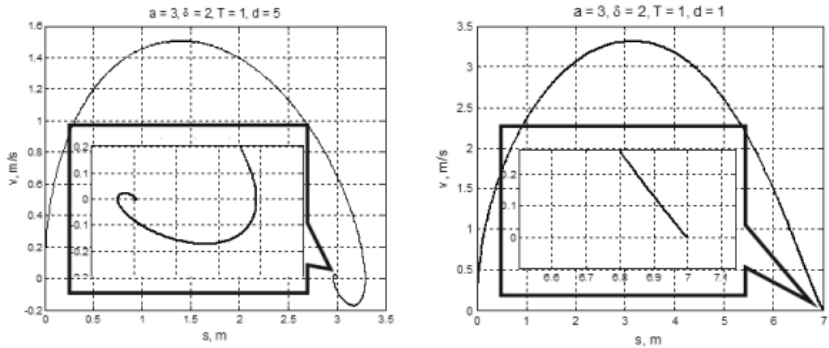}
\caption{Phase diagramm 'velocity vs. distance'. Oscillations of the velocity and the gap to the fixed leader (left), and non-oscillating solution due to conditions obtained (right).}
\label{fig:2}       
\end{figure}
Fig.~\ref{fig:3} demonstrates the velocity series in case of vehicle platoon. We examine 9 vehicles moving one by one in accordance with the model (\ref{eq:02}). The 10th car is fixed. In the first case the stability condition is fulfilled for all cars and no oscillating solution is observed. Then we break stability condition for the 5th vehicle and observe velocity oscillations of this and all subsequent ones.
\begin{figure}[h]
\includegraphics[width = 1\textwidth]{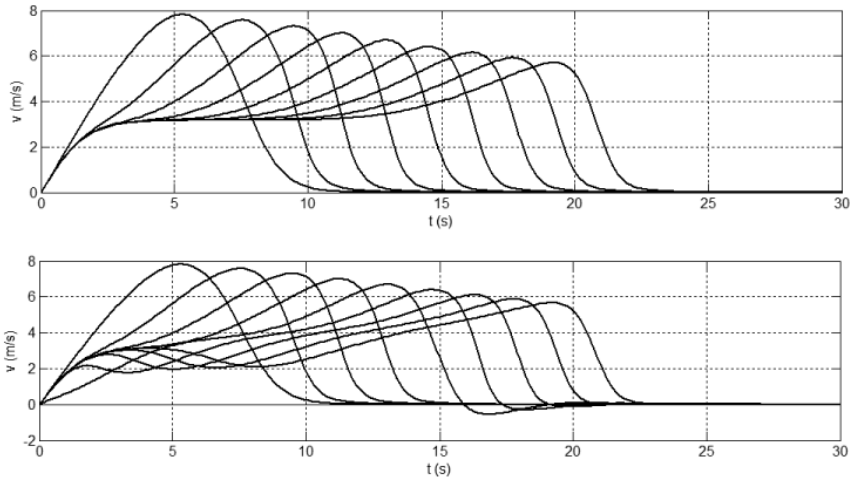}
\caption{Platoon stability - 9 vehicles approaching the 10th fixed car. Top: model parameters for every vehicle - $a = 2, T = 1.1, s' = 1$. Stability conditions are satisfied for every one - no oscillating solution. Below: model parameters for the 5th vehicle - $a = 2, T = 1.1, s' = 4$, for others are the same as in the first case. Stability conditions are broken for the 5th one - oscillations are observed.}
\label{fig:3}       
\end{figure}
}

\section{Model calibration}
\label{sec:4}
{The calibration procedure can be carried out on two different levels that are macroscopic and microscopic ones. This former implements calibration with respect to macroscopic traffic data, for example, flow-density data for the specific region, or estimates origin-destination matrices \cite{Cascetta}. The latter treats vehicles as an individual entities and uses microscopic trajectory data \cite{Kesting2008, Ossen}. The goal is to determine the optimal model parameters that better reproduce vehicle dynamics (acceleration on free-road, deceleration process) and drivers' behaivior (distance between vehicles in a steady-state flow). Consequently, the calibration is evaluated on the basis of intra-driver and inter-driver criteria. \newline\indent
Our framework is intended for microlevel calibration and can be used for any parametrical traffic model that is described by ODE or DDE \cite{Li}. Let the vehicle's dynamics is defined by the ODE with following initial conditions
$$
\begin{array}{l}
  \dfrac{dh}{dt} = v_L - v
	\\ [4mm]
	\dfrac{dv}{dt} = A(v_L,v,\theta) 
	\\ [4mm]
   h(t_0) = h_0, v(t_0)=v_0
\end{array}
$$
Here $\theta$ contains the set of the model parameters to be determined. In case of the first modification $\theta = (a,\delta,T,c)$. Also the real data - velocity time history ${\left \{ \tilde{v_i} \right \}}_{i=1}^N$ and trajectory data ${\left \{ \tilde{h_i} \right \}}_{i=1}^N$ are provided. We have to determine the model parameters that minimize the objective function with respect to vehicle's speed and its position. The main difficulty is that the second one can not be obtained analytically. Using the Euler method for numerical integration we formulate optimization problem with constraints as below
\begin{equation}
\left\{ 
  \begin{array}{l l}
    w_1||v-\tilde{v}||^2+w_2||h-\tilde{h}||^2 \to min \\
    s_{i+1} - s_i = \Delta t \cdot v_i, v_{i+1} - v_i = \Delta t \cdot A(v_i, v_{Li}, h_i, \theta) \\
		h(t_0) = h_0, v(t_0)=v_0 \\
		\text{+ local stability condition (\ref{eq:09})}
  \end{array} \right.
\label{eq:03}
\end{equation}
The no oscillation condition obtained for the first modification is here incorporated in the optimization problem as an inequality constraint. \newline\indent
We consider two sequent problems for calibrating procedure:
\begin{itemize}
	\item {free-road acceleration (no leader presents),}
	\item {the emergency deceleration to avoid collision.} 
\end{itemize}
Solution of the first optimization problem contains values of two parameters - $a$ and $\delta$. Then solving the second one we obtain values for last two parameters – $T$ and $c$.
\begin{figure}[h]
\begin{center}
\includegraphics[width = 0.9\textwidth]{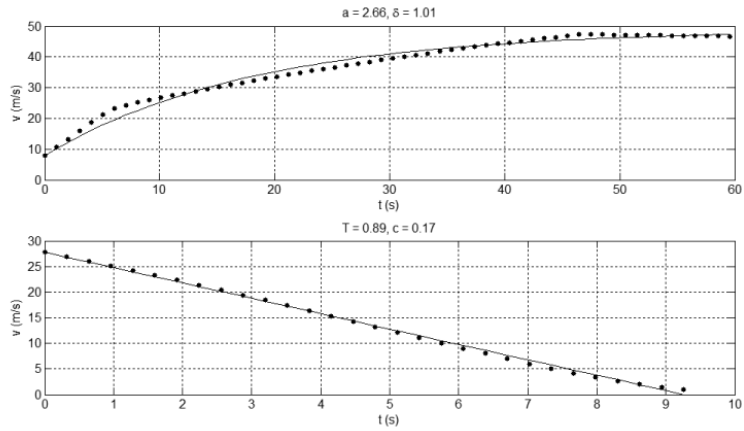}
\end{center}
\caption{Solution of the two-step calibrating problem - comparison of simulated and empirical data. The numerical solution and real data are shown as solid and dashed curves respectively. Top: free-road acceleration case and values for $a$ and $\delta$. Below: emergency deceleration case and values for $T$ and $c$.} 
\label{fig:4}       
\end{figure}
This calibrating procedure was evaluated for several vehicle models. In summary, this framework provides the parameter space that provides real dynamics of the specific car model and simultaneously ensures solution stability.
}

\section{Discussions}
\label{sec:5}
{Driving simulator is an effective tool as a training system. It allows to put trainee in real environment with road infrostructure (signs, road surface markings, traffic lights), vehicle traffic and pedestrians. The adequacy of environment reprodution affects on driver's perception and feelings and defines the quality of learning process. 
The realistic microscopic models are required for vehicle traffic simulation. The training process implies different environment conditions. Moreover, traffic flow should be diverse from two points of view - drivers' behaivior and cars' dynamics. In other words, the mathematical models used should reproduce dynamics of the specific car model, drivers' behaviour and take into account weather conditions (e.g. day/night time, fog, snow, surface icing). 

In this paper we propose two microscopic models for using with driving simulators. The first one is developed on the basis of the well-known Intelligent Driver Model (IDM). Deriving a continious weight-function we combine two different modes - free-road acceleration and interaction with the preceeding car - in one continiously differentiable acceleration function. This model ensures any predefined distance in a steady state and, thus, allows taking into account weather conditions and consider different phycological types of drivers. This model is investigated with linear stability analysis for the case of vehicle platoon. The stability conditions are obtained in general form and assume that every vehicle in a platoon has its own continious acceleration function.

To consider human reaction time explicitly we propose the second microscopic model. In accordance with the acceleration function drivers response to their headway via the delay $\tau$. This model combines the first modification and the optimal velocity. When compated to the IDM this model is not crash-free, always results in real acceleration values and allows to simulate collisions. Moreover, to guarantee safe stop before fixed leader with different initial conditions for velocity and position we modified optimal velocity function and calibrated it to fulfill this objective.

The model calibrating procedure is an essential part of model preprocessing. In this work we construct the framework for microlevel calibration. As inputs the velocity and position time history are used. We consider two scenaria - acceleration to the maximum speed with no leader presented and the emergency deceleration with the constrant deceleration rate. The minimization of the objective function is evaluated both with respect to the vehicle's velocity and position, each of which has its own weight coefficient. This way, we can indicate what is more important - vehicle's gap or its speed agreement. This framework can be used for any microscopic model with parameters to be determined on the basis of ODEs or DDEs. In this paper, we demonstrate the results for the first microscopic model presented.

For the further work, we are planning to investigate the presented time-delay model with stability analysis. Some articles \cite{Orozs, Orozs2009} and approaches \cite{Lakshmanan} have already been studied on this subject. The goal is to find out weather this model admits real human reaction times and simultaneously ensures solution stability. Another issue we are interested in is the effective numerical schemes. The Euler method used now does not work well for such time steps as human reaction time. As a result, we need to use more robust scheme in calculations. Moreover, this scheme should not be numerically consuming in order to simulate city-scale traffic in a real time.}

%
%
%

\end{document}